\documentclass[%
 reprint,
superscriptaddress,
 amsmath,amssymb,
 aps,
 prl,
floatfix,
]{revtex4-2}

\usepackage{graphicx}
\usepackage{dcolumn}
\usepackage{bm}
\usepackage{lineno}

\usepackage{epsfig,color}
\usepackage{soul}
\usepackage{amssymb}

\begin{document}
\preprint{APS/123-QED}

\title{Size-Dependent Properties of Miura-ori Tessellations}

\author{C. Baek}
\affiliation{Department of Mechanical Engineering, Seoul National University, Seoul 08826, Republic of Korea}
\author{T. Tachi}
\affiliation{Department of General Systems Studies, Graduate School of Arts and Sciences, The University of Tokyo, Meguro-ku, Tokyo 153-8902, Japan}
\author{J. Yang}
\thanks{Corresponding author. Email: jkyang11@snu.ac.kr}
\affiliation{Department of Mechanical Engineering, Seoul National University, Seoul 08826, Republic of Korea}
\author{H. Yasuda}
\thanks{Corresponding author. Email: yasuda.hiromi@jaxa.jp}
\affiliation{Institute of Space and Astronautical Science, Japan Aerospace Exploration Agency, Sagamihara, Kanagawa 252-5210, Japan}
\affiliation{Department of Aeronautics and Astronautics, University of Tokyo, Bunkyo-ku, Tokyo 113-8656, Japan}


\date{\today}

\begin{abstract}
We investigate the size-dependent behavior of Miura-ori-based origami tessellations by changing the number of origami unit cells.
For large tessellations, the Miura-ori sheet generally exhibits a negative in-plane Poisson's ratio, whereas if the size of the Miura-ori tessellations becomes small, the transition between positive and negative Poisson's ratio emerges in the middle of the folding process.
Here, we show that such a transitioning point, i.e., zero Poisson's ratio, yields a kinematic locking state.
We also experimentally demonstrate the tunable locking behavior altered by tessellation sizes.
Extending the analysis to three-dimensional origami tessellations, we find that the direction of kinematic locking changes depending on the tessellation size.
Varying tessellation size thus enables control over both the onset and the direction of locking in origami metamaterials.
\end{abstract}

\maketitle

Size-dependent mechanical properties arise across materials science. Finite dimensions often constrain plastic flow and strengthen metals~\cite{uchic2004sample, fleck1994strain, stolken1998microbend}, while nominal strength and fracture depend on specimen size in composites~\cite{bazant1996size, bazant2019fracture, ko2019effect}.
Beyond conventional materials, architected metamaterials exhibit analogous size effects arising from geometry: stiffness and deformation depend on metrics such as a characteristic length or the number of unit cells. In many such systems, deformation localizes near free boundaries, causing bulk homogenization to fail when the structure comprises only a few unit cells~\cite{frenzel2017three, Coulais2018, Yang2021}.

Origami-based metamaterials, especially those stemming from rigid-foldable patterns such as the Miura-ori~\cite{koryo1985method}, provide a tractable platform for examining such geometry-property relations, as their mechanical response can often be expressed directly in geometric terms~\cite{Dudek2025, Schenk2013_PNAS, wei2013geometric, yasuda2015reentrant}.
The Miura-ori possesses a single global degree of freedom (DOF) that persists across tessellation sizes, and its boundaries introduce no additional freedom, unlike systems where edge-localized modes drive finite-size effects~\cite{imada2025maxwell, chen2016topological}.
Of particular interest in engineering applications of origami is controlling the available DOFs to achieve structures that are foldable yet kinematically lockable.
Prior studies typically induce locking through geometric heterogeneity or defects~\cite{Schenk2013_PNAS, silverberg2014using},
imposed symmetry~\cite{doi:10.1098/rspa.2023.0956}, or panel self-contact~\cite{doi:10.1098/rspa.2016.0682, GAO2022108806}.
However, these approaches often restrict the accessible folding range and introduce localized effects, such as non-periodic mechanics or facet compliance beyond rigid-origami kinematics, complicating geometric descriptions of locking.

In this Letter, we demonstrate the onset of size-dependent, intrinsic locking in defect-free origami tessellations based on the Miura-ori.
We show that the Poisson's ratio can switch between positive and negative values not only with the geometric parameters but also with the number of unit cells. At the zero-Poisson's-ratio transition, the structure exhibits kinematic locking.
Although such kinematic locking in origami structures has been studied previously~\cite{tachi2012rigid, cheung2014origami, 2019HiromiTMP}, the size effect of origami tessellations remains largely unexplored.
Extending this finding to three-dimensional Miura-ori tube architectures reveals a broader mechanism: a quadratic form incorporating size-dependent Poisson's ratios governs directional locking across all tessellation scales and defines the locking-direction surface as an elliptic cone. Interestingly, for small tessellations, the cone's symmetry axis switches mid-folding; at the transition, two Poisson's ratios vanish, yielding anomalous locking directions described by intersecting planes.

\begin{figure}[htb]
    \includegraphics[width=0.49\textwidth]{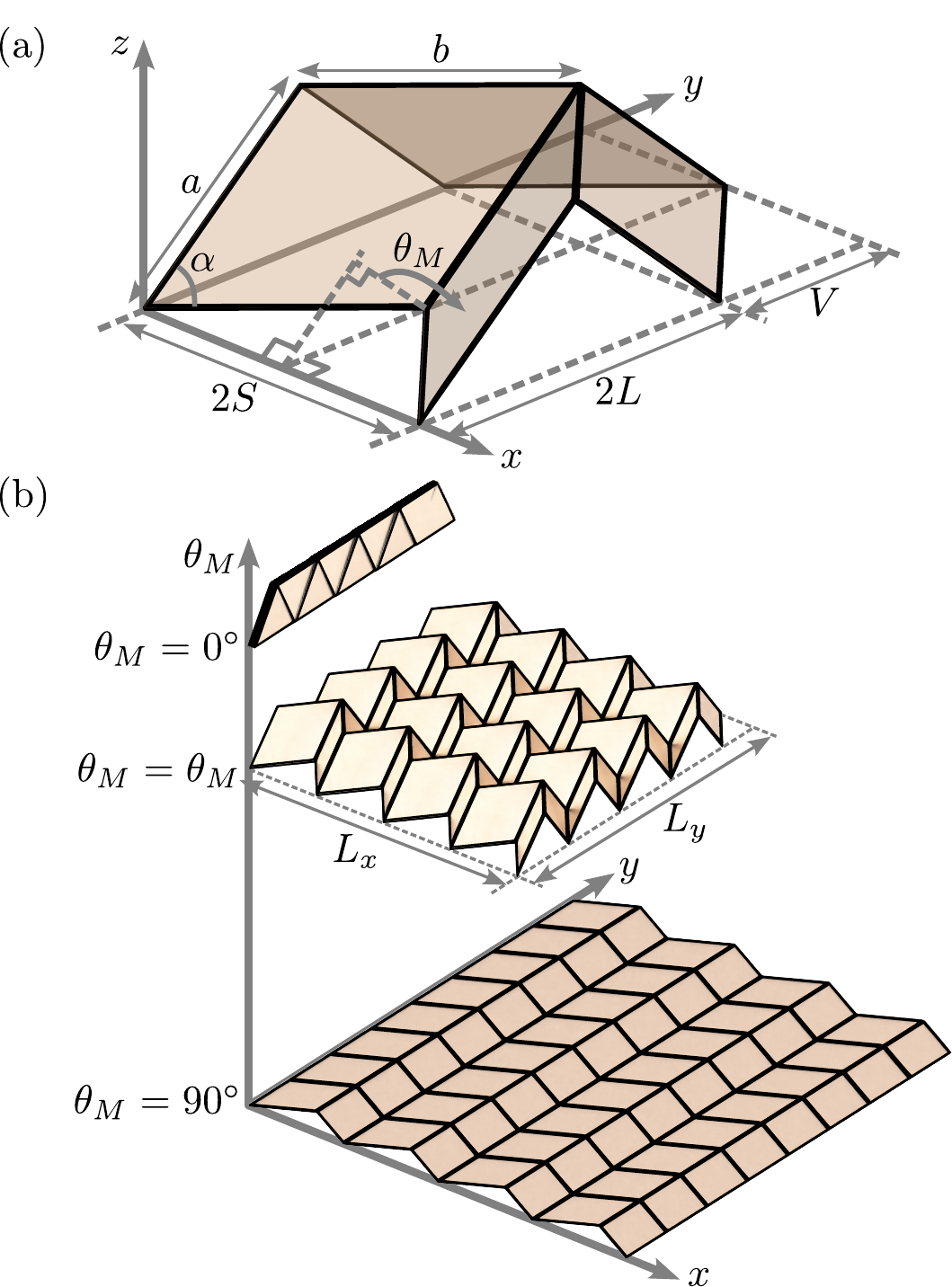}
    \caption{Geometry of the Miura-ori. (a) Miura-ori unit cell. 
    (b) Folding motion of $4 \times 4$ Miura-ori whose overall width and breadth are denoted by $L_x$ and $L_y$, respectively.
    }\label{fig:geometry}
\end{figure}

We begin by defining the geometry of the Miura-ori.
Figure~\ref{fig:geometry}(a) shows the unit cell of the Miura-ori, which consists of four identical parallelogram facets characterized by two side lengths, $a$ and $b$, and an interior angle $\alpha$. 
To describe the folded configuration, we use the folding parameter $\lambda=\sin \alpha \sin \theta_M$, where $\theta_M$ is the folding angle as shown in Fig.~\ref{fig:geometry}(a).
Here, we define key dimensions of our unit cell ($S$, $L$, and $V$), expressed as
\begin{equation}\label{eq:unitcell_dimensions}
    S=b\lambda, \quad
    L=\frac{a \cos \alpha}{\sqrt{1-\lambda^2}}, \quad
    V=b \sqrt{1-\lambda^2}.
\end{equation}

To investigate the kinematics of the Miura-ori with different sizes (i.e., size effect),  we consider the Miura-ori sheets composed of $N_x \times N_y$ unit cells [see Fig.~ \ref{fig:geometry}(b) for $N_x=N_y=4$ case].
The overall dimensions $L_x$ and $L_y$ are expressed in terms of the unit-cell parameters [Fig.~\ref{fig:geometry}(b)]:
\begin{align}
    \begin{split}
        L_x &= 2N_x S,
    \end{split}\label{eq:L_x}\\
    \begin{split}
        L_y &=  2 N_y L + V.
    \end{split}\label{eq:L_y}
\end{align}
Then, we obtain the in-plane infinitesimal Poisson's ratio defined by $\nu_{xy}=-(dL_y/L_y)/(dL_x/L_x)$, which gives
\begin{equation}\label{eq:nu_xy}
    \nu_{xy} = -\frac{\cos \alpha - (b/a) (1-\lambda^2)/(2N_y)}{\cos \alpha + (b/a) (1-\lambda^2) /(2N_y)} \frac{\lambda^2}{1-\lambda^2}.
\end{equation}
Here, if we consider the bulk property when $N_y \to \infty$, the Poisson's ratio becomes $\nu_{xy}=-\lambda^2/(1-\lambda^2)$ ~\cite{Schenk2013_PNAS}.

\begin{figure}[tb]
    \includegraphics[width=0.5\textwidth]{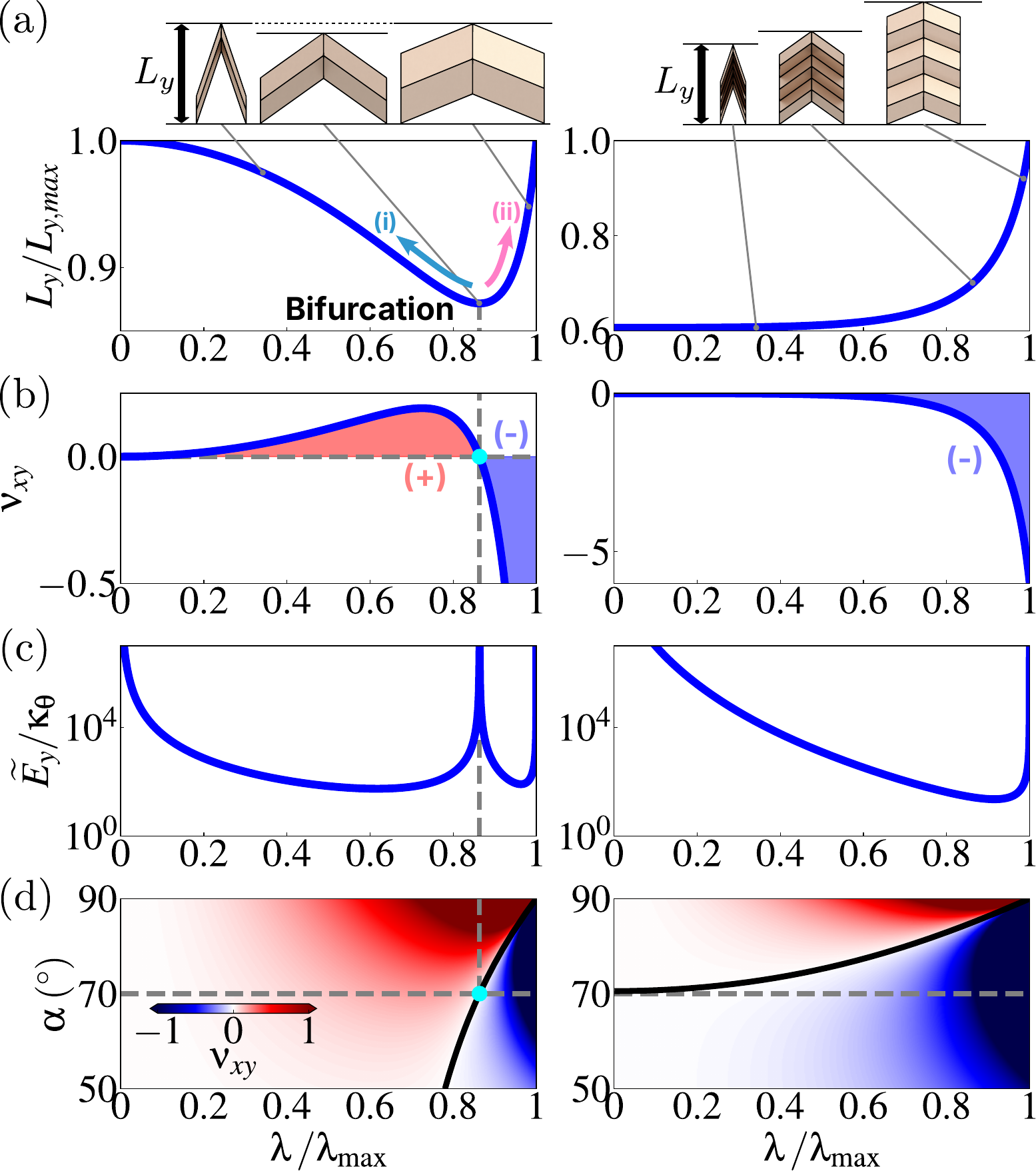}
    \caption{Size-dependent behavior of the Miura-ori. We compare two Miura-ori tessellations with different sizes in the $y$ direction: $N_y=1$ (left) and $N_y=3$ (right). For both cases, $N_x=1$ and $(b/a, \alpha)=(2, 70^\circ)$.  (a) Normalized length $L_y/L_{y, \mathrm{max}}$ versus $\lambda/\lambda_{\mathrm{max}}$; insets show corresponding deployment sequences.
    (b) Poisson's ratio $\nu_{xy}$ versus $\lambda/\lambda_{\mathrm{max}}$. Red: positive, Blue: negative. (c) Nondimensionalized stiffness $\widetilde{E}_y/k_\theta$ versus $\lambda/\lambda_{\mathrm{max}}$.
    (d) Colormaps of $\nu_{xy}$ in the $\alpha-\lambda/\lambda_{\mathrm{max}}$ plane; the black solid curve marks the locking trajectory.
    }
    \label{fig:Locking}
\end{figure}

Figure~\ref{fig:Locking} illustrates how the tessellation size $N_y$ governs the Miura-ori kinematics. Note that throughout this study, we fix the unit-cell geometry ($\alpha=70^\circ$, $b/a=2$) for all analyses.
Figure~\ref{fig:Locking}(a) shows the structural length $L_y$ against $\lambda/\lambda_{\textrm{max}}$ for $N_y=1$ (left) and $N_y=3$ (right).
Here, $\lambda_{\textrm{max}}=\sin \alpha$.
Our analysis reveals that the Miura-ori with $N_y=1$ exhibits a non-monotonic shape change, whereas $L_y$ varies monotonically for $N_y=3$.
For $N_y=1$, $L_y$ exhibits an interior minimum, which serves as a compression-stable equilibrium.
Then, if the structure at such a minimum point is put under tension, the equilibrium is unstable and the motion bifurcates toward either the fully folded ($\lambda/\lambda_{\textrm{max}}=0$) or the fully deployed ($\lambda/\lambda_{\textrm{max}}=1$) state, depending on the perturbation direction [paths (i) and (ii) in Fig.~\ref{fig:Locking}(a, left)].
In contrast, for $N_y=3$, no bifurcation occurs.
This size dependence follows from Eq.~(\ref{eq:L_y}): the tip length ($V$) is set solely by the unit-cell geometry [Fig.~\ref{fig:geometry}(a)], while the body part $2N_y L$ scales with $N_y$.
These contributions vary oppositely with folding, and create an interior minimum in $L_y$ for small $N_y$. As $N_y$ increases, $V$ becomes negligible and $L_y$ becomes monotonic (see Supplemental Material~\cite{supp}).

We also examine the Poisson's ratio $\nu_{xy}$ [Fig.~\ref{fig:Locking}(b)].
For $N_y=1$, $\nu_{xy}$ is initially positive and becomes negative during unfolding.
The sign change coincides with the interior minimum of $L_y$, as $L_x$ is monotonic and $\nu_{xy}=-(dL_y/L_y)/(dL_x/L_x)$.
In contrast, for $N_y=3$, $\nu_{xy}$ remains negative throughout the entire folding range~\cite{Schenk2013_PNAS}, consistent with the monotonic $L_y$ and the absence of an interior extremum.

The Poisson's ratio can be linked to directional stiffness of the structure through the rigid-origami framework, where facets are modeled as rigid panels and creases as linear torsional springs with torsional stiffness per unit crease length, $\kappa_\theta$.
We introduce the nondimensionalized axial stiffnesses $\widetilde{E}_x$ and $\widetilde{E}_y$, normalized by the cross-sectional area and structural length along the $x$ and $y$ axes, respectively. 
Since the Miura-ori is a one-degree-of-freedom mechanism, its in-plane strains depend on a single folding parameter, imposing the reciprocal symmetry $\nu_{xy} = 1/\nu_{yx}$ \cite{liu2022triclinic}. Together with the orthotropic reciprocity condition $\nu_{xy}/\widetilde{E}_x=\nu_{yx}/\widetilde{E}_y$, this yields
\begin{equation}\label{eq:E_poisson}
    \widetilde{E}_y = \frac{\widetilde{E}_x}{\nu_{xy}^2}.
\end{equation}
Notice that $\widetilde{E}_y$ is not independent of $\widetilde{E}_x$; in this 1-DOF rigid-origami framework, all in-plane stiffnesses are kinematically coupled.

Figure~\ref{fig:Locking}(c) shows $\widetilde{E}_y/\kappa_\theta$.
For $N_y=1$, $\widetilde{E}_y$ diverges at the mid-deployment state when $\nu_{xy}=0$, as predicted by Fig.~\ref{fig:Locking}(a, b) and Eq.~(\ref{eq:E_poisson}).
From Eq.~(\ref{eq:nu_xy}), the locking occurs when
\begin{equation}\label{eq:condition2}
    \lambda = \sqrt{1-\frac{2N_y}{(b/a)} \cos \alpha}.
\end{equation}
This locking condition is physically admissible when $0\leq\lambda \leq \sin \alpha$, equivalent to
\begin{equation}\label{eq:inequality}
    \cos^{-1}{\left[\min \left(\frac{(b/a)}{2 N_y}, \frac{2N_y}{(b/a)}\right) \right]} \leq \alpha \leq \frac{\pi}{2}.
\end{equation}
Thus, the unit-cell geometry ($\alpha$, $b/a$) and the tessellation size $N_y$ jointly determine the locking behavior of the structure.

The size dependence is also evident in the colormaps of $\nu_{xy}$ against $\alpha$ and $\lambda/\lambda_{\textrm{max}}$ [Fig.~\ref{fig:Locking}(d)].
The solid black curve indicates $\nu_{xy}=0$, i.e., the locking trajectory. For $N_y = 1$, a horizontal line at $\alpha=70^\circ$ intersects this curve at $\lambda/\lambda_{\textrm{max}}=0.86$, which coincides exactly with the stiffness spike in Fig.~\ref{fig:Locking}(c, left), but not for $N_y = 3$ (see Fig.~\ref{fig:Locking}(c, right)).

\begin{figure}[tb]
    \includegraphics[width=0.5\textwidth]{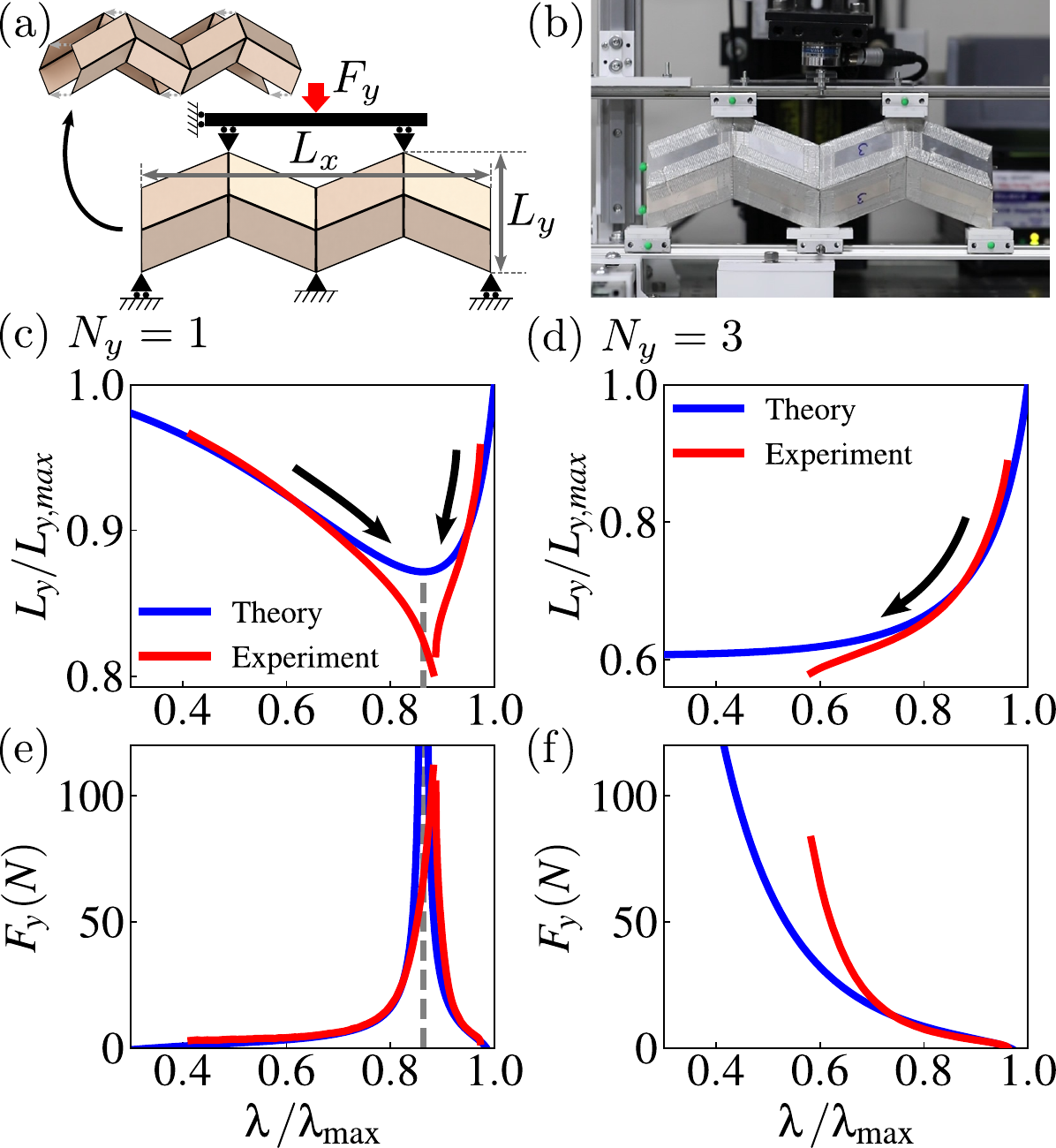}
    \caption{Experimental validation of size-dependent locking in Miura-ori under uniaxial compression. (a) Mechanical boundary conditions. Inset: Miura-ori tube fabrication from two identical Miura-ori sheets. (b) Photo of the experiment. (c),(d) Comparison of theoretical and experimental mean normalized length $L_y/L_{y, max}$ versus $\lambda/\lambda_{\mathrm{max}}$ for $N_y=1, 3$; black arrows indicate the compression direction. (e),(f) Corresponding theoretical and experimental mean $F_y$-$\lambda/\lambda_{\mathrm{max}}$ curves. Experimental curves show the mean of five trials; standard deviations are typically no larger than the line width and omitted for clarity. All samples share identical geometry with $\alpha=70^\circ$, $b/a=2$, and $N_x=2$.
    }
    \label{fig:Experiment1}
\end{figure}

To experimentally verify the locking behavior arising from the size effect, we fabricate prototypes and conduct uniaxial compression experiments.
Miura-ori tubes are fabricated by joining two identical Miura-ori sheets made of 0.3 mm-thick SUS304H stainless-steel facets and fiber-reinforced tape hinges. 
All specimens share the identical unit-cell geometry.
Figures~\ref{fig:Experiment1}(a) and (b) show a schematic and a photograph of the experimental setup, respectively.
The force $F_y$ is measured during compression using a load cell, while the in-plane kinematics ($L_x$, $L_y$) are tracked by digital image correlation.
Also, a fitted rigid-origami model is used to compare our prediction with the experimental results. 
For the Miura-ori with $N_y=1$ exhibiting kinematic bifurcation,
we perform compression tests starting from two initial configurations: one from the unfolded side of the locking state and one from the folded side, as indicated by the arrows in Fig.~\ref{fig:Experiment1}(c).
For the $N_y=3$ specimen, the compression is applied only from the unfolded configuration, as shown in Fig.~\ref{fig:Experiment1}(d). See the Supplemental Material \cite{supp} for fabrication and fitting details; see Supplemental Movie S1 for uniaxial compression tests.

Figures~\ref{fig:Experiment1}(c) and (d) show the experimentally measured $L_y$ and the theoretical prediction plotted against $\lambda/\lambda_{\textrm{max}}$ for $N_y=1$ and $N_y=3$, respectively.
Figures~\ref{fig:Experiment1}(e) and (f) present the measured and analytical axial force $F_y$ for each case.
For $N_y=1$, both theory and experiment exhibit a non-monotonic $L_y$, producing a sharp spike in $F_y$ near $\lambda/\lambda_{\textrm{max}}=0.86$ obtained from Eq.~(\ref{eq:condition2})~[dashed line in Fig. \ref{fig:Experiment1}(d)].
In contrast, for $N_y=3$, $L_y$ remains monotonic and compression drives the structure toward the fully folded endpoint, accompanied by a monotonic increase of $F_y$ over the measured range.
Despite facet deformation which allows $L_y$ to drop below rigid-origami prediction and causes the force response to deviate at large compression, the overall trends confirm the size-dependent onset of a mid-deployment locking state.

\begin{figure*}[htb]
\includegraphics[width=1.0\linewidth]{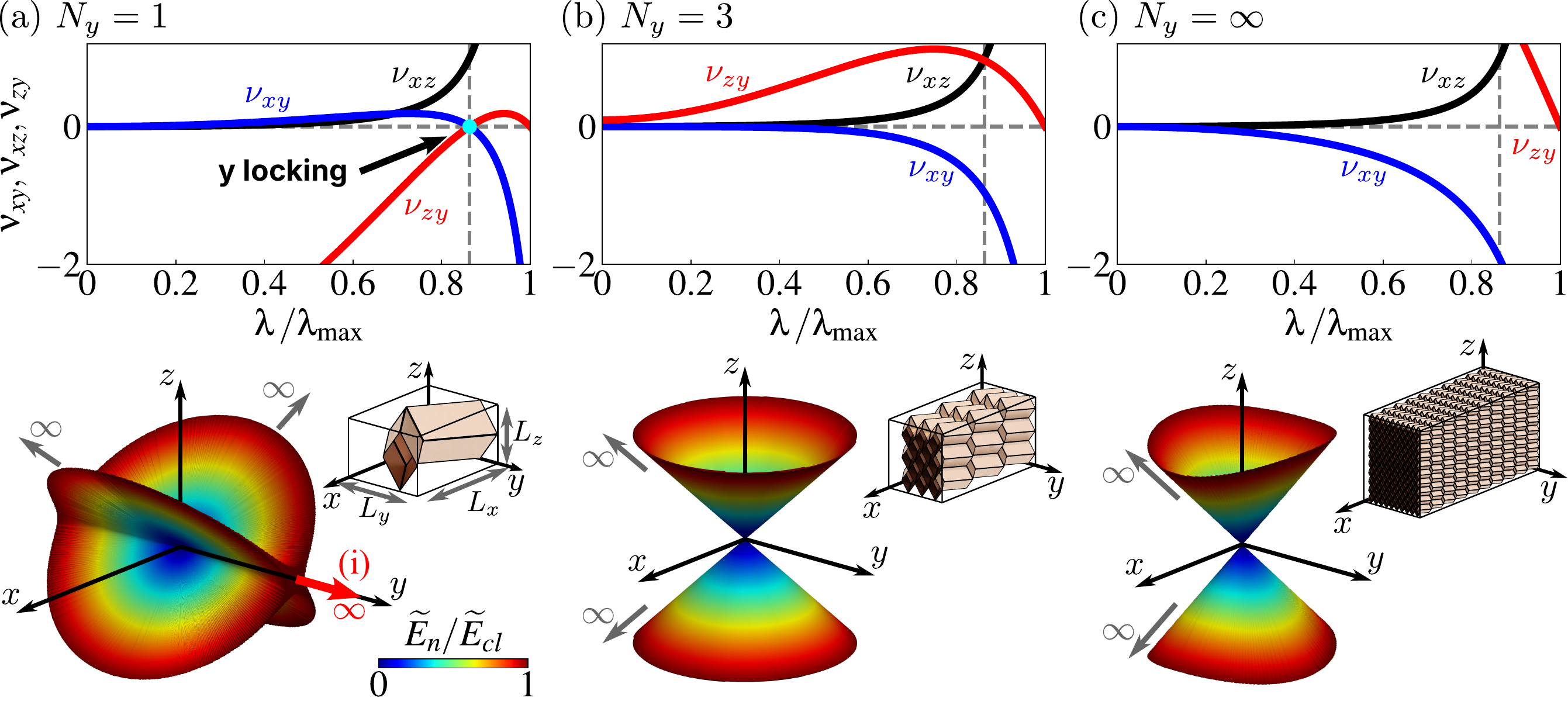}
\caption{Poisson's ratios and nondimensionalized stiffness of Miura-ori structures with different numbers of unit cells $N_y$ and identical unit-cell geometry ($\alpha=70^\circ$, $b/a=2$). Poisson's ratios $\nu_{xy}$, $\nu_{xz}$, $\nu_{zy}$ are plotted as functions of $\lambda/\lambda_{\textrm{max}}$.
Insets show the nondimensionalized stiffness distributions $\widetilde{E}_n$ (diverging stiffness clipped at $\widetilde{E}_{cl}$ for visualization) and the rendered structures at $\lambda/\lambda_{\textrm{max}}=0.86$; color intensity indicates stiffness magnitude, and gray arrows mark representative directions on the divergence surfaces of $\widetilde{E}_n$. All insets share a common coordinate system. (a) $N_y=1$; arrow (i) indicates locking along the $y$ axis. (b) $N_y=3$; (c) $N_y\to\infty$. 
}
\label{fig:3DElastic}
\end{figure*}

The relation between Poisson's ratio and stiffness in the Miura-ori sheet suggests a broader principle: mid-deployment kinematic locking occurs when Poisson's ratio becomes zero.
We further explore this principle in a three-dimensional Miura-ori tube by evaluating its response in any arbitrary direction.
The 3D structure is formed by stacking identical Miura-ori sheets alternately flipped along the normal direction.
The Poisson's ratios of the Miura-ori tube are $\nu_{xy}$ expressed by Eq.~(\ref{eq:nu_xy}),
\begin{equation}\label{eq:nu_xz}
        \nu_{xz} = \frac{\cos^2 \alpha}{\sin^2 \alpha - \lambda^2}\frac{\lambda^2}{1-\lambda^2},
\end{equation}
and $\nu_{zy}=-\nu_{xy}/\nu_{xz}$~(see Supplemental Material \cite{supp}).

The nondimensionalized stiffness $\widetilde{E}_n$ along unit vector $n=(n_x, n_y, n_z)$ can be written solely using the kinematic properties and $\widetilde{E}_x$~\cite{supp}:
\begin{equation}\label{eq:Elastic_Modulus_simp}
    \widetilde{E}_n = \frac{\widetilde{E}_x}{\left(n_x^2 -\nu_{xy}n_y^2 -\nu_{xz}n_z^2\right)^2}.
\end{equation}
This formulation provides a scale-independent measure of directional rigidity and recovers the conventional elastic modulus form in the large tessellation continuum limit. 
Also, Eq.~(\ref{eq:Elastic_Modulus_simp}) diverges, leading to kinematical locking, when
\begin{equation}\label{eq:quadratic_form}
    n_x^2-\nu_{xy} n_y^2 -\nu_{xz} n_z^2=0.
\end{equation}
It is noteworthy that the locking behavior of the Miura-ori tube can be characterized by a quadratic surface, and one of its coefficients, $\nu_{xy}$, varies with tessellation size.

Figure~\ref{fig:3DElastic} presents the Poisson's ratios as functions of $\lambda/\lambda_{\textrm{max}}$ for $N_y=$ 1, 3, and $\infty$, with identical unit-cell geometry. 
For $N_y=1$, $\nu_{xz}$ is always positive in the entire folding range, whereas both $\nu_{xy}$ and $\nu_{zy}$ cross zero at $\lambda/\lambda_{\textrm{max}}=0.86$~[Fig.~\ref{fig:3DElastic}(a)].
Here, the quadratic condition (Eq.~(\ref{eq:quadratic_form})) defines an elliptic cone of locking directions, with its axis of symmetry along $z$ ($x$) for $\nu_{xy}<0$ ($\nu_{xy}>0$) (see Supplemental Material~\cite{supp} and Supplemental Movie S2 for the evolution of $\widetilde{E}_n$ as $\nu_{xy}$ changes sign).
Interestingly, if $\nu_{xy}$ becomes zero, the cone degenerates into two planes $n_x=\pm\sqrt{\nu_{xz}}n_z$ intersecting along the $y$ axis.
The three-dimensional stiffness map at this point, computed from Eq.~(\ref{eq:Elastic_Modulus_simp}), diverges on the two predicted planes, with the $y$-directional locking highlighted by arrow~(i) in Fig.~\ref{fig:3DElastic}(a).
Although the Miura-ori sheet also locks in the $y$-direction (e.g., Fig.~\ref{fig:Locking}), its derivative (3D) tessellation exhibits richer locking capability in three-dimensional space.

In contrast, for $N_y=3$ and $N_y \rightarrow \infty$ [Fig.~\ref{fig:3DElastic}(b), (c)], $\nu_{xy}$ and $\nu_{zy}$ do not cross zero mid-deployment.
The stiffness maps at the same $\lambda/\lambda_{\textrm{max}}$ as in panel (a) are shown below for comparison. They show that $\widetilde{E}_n$ diverges on an elliptic cone without $y$-directional locking. Thus, we confirm that varying tessellation size selects the locking-direction surface and thereby activates or suppresses $y$-directional locking, demonstrating size-controlled programmability in origami-based structures. See Supplemental Movie S3~\cite{supp}, which extends Fig.~\ref{fig:3DElastic} by showing $\widetilde{E}_n$ at several $\lambda/\lambda_{\textrm{max}}$ values for $N_y=1 \text{--} 10$.

In summary, we have investigated the size-dependent mechanical response of the origami tessellations based on the Miura-ori within the rigid-origami framework.
For small tessellations, finite-size effects create a local minimum in transverse width, resulting in zero Poisson's ratio.
At this singular state, the axial stiffness diverges, yielding a mid-deployment locking state.
On the other hand, as tessellation size increases, such locking behavior disappears.
We further develop our analysis to three-dimensional Miura-ori tubes, where kinematic locking is characterized by a quadratic surface taking the shape of an elliptic cone. The size-dependent sign change of the Poisson's ratio reorients this locking surface; moreover, a zero crossing triggers a degeneracy into two planes, thereby activating locking along a principal axis. Thus, the locking of small Miura-ori sheets emerges as a special case of a broader Poisson-governed locking condition intrinsic to Miura-ori tubes.
These results establish a unified geometric framework linking scale, kinematics, and anisotropy, enabling size-controlled programming of directional stiffness and locking.

\textit{Acknowledgments}---We thank Yasuhiro Miyazawa for the discussions on experiments, Sunghyun Kim and Yubin Oh for help with metal fabrication.
H.Y. gratefully acknowledges support
from the JSPS KAKENHI JP22K18750.
C.B. and J.Y. acknowledge the support from Air Force Office of Scientific Research (FA2386-24-1-4051), SNU-IAMD, SNU-IOER, and National Research Foundation grants funded by the Korean government [2023R1A2C2003705 and 2022H1D3A2A03096579].

\textit{Data Availability}---The data are not publicly available. The data are available from the authors upon reasonable request.

\nocite{Schenk2013_PNAS, liu2022triclinic, opencv_library, yasuda2015reentrant}

\bibliography{apssamp}

@PREAMBLE{
 "\providecommand{\noopsort}[1]{}" 
 # "\providecommand{\singleletter}[1]{#1}%" 
}

@article{uchic2004sample,
  title={Sample dimensions influence strength and crystal plasticity},
  author={Uchic, Michael D and Dimiduk, Dennis M and Florando, Jeffrey N and Nix, William D},
  journal={Science},
  volume={305},
  number={5686},
  pages={986--989},
  year={2004},
  publisher={American Association for the Advancement of Science}
}

@article{bazant1996size,
  author = {Ba\v{z}ant, Zdeneˇk P. and Daniel, Isaac M. and Li, Zhengzhi},
    title = {Size Effect and Fracture Characteristics of Composite Laminates},
    journal = {Journal of Engineering Materials and Technology},
    volume = {118},
    number = {3},
    pages = {317-324},
    year = {1996},
    month = {07},
    issn = {0094-4289},
    doi = {10.1115/1.2806812},
}

@book{bazant2019fracture,
  title={Fracture and size effect in concrete and other quasibrittle materials},
  author={Ba\v{z}ant, Zdenek P and Planas, Jaime},
  year={2019},
  publisher={Routledge}
}

@article{fleck1994strain,
  title={Strain gradient plasticity: theory and experiment},
  author={Fleck, NA and Muller, GM and Ashby, Mike F and Hutchinson, John W},
  journal={Acta Metallurgica et materialia},
  volume={42},
  number={2},
  pages={475--487},
  year={1994},
  publisher={Elsevier}
}

@article{stolken1998microbend,
  title={A microbend test method for measuring the plasticity length scale},
  author={St{\"o}lken, J Sꎬ and Evans, AG},
  journal={Acta Materialia},
  volume={46},
  number={14},
  pages={5109--5115},
  year={1998},
  publisher={Elsevier}
}

@article{ko2019effect,
  title={Effect of the platelet size on the fracturing behavior and size effect of discontinuous fiber composite structures},
  author={Ko, Seunghyun and Yang, Jinkyu and Tuttle, Mark E and Salviato, Marco},
  journal={Composite Structures},
  volume={227},
  pages={111245},
  year={2019},
  publisher={Elsevier}
}

@article{imada2025maxwell,
  title={Maxwell origami tube},
  author={Imada, Rinki and Tachi, Tomohiro},
  journal={Physical Review Research},
  volume={7},
  number={1},
  pages={013032},
  year={2025},
  publisher={APS}
}

@article{doi:10.1098/rspa.2016.0682,
  title={Self-locking degree-4 vertex origami structures},
  author={Fang, Hongbin and Li, Suyi and Wang, KW},
  journal={Proceedings of the Royal Society A: Mathematical, Physical and Engineering Sciences},
  volume={472},
  number={2195},
  pages={20160682},
  year={2016},
  publisher={The Royal Society}
}

@article{GAO2022108806,
title = {Origami-inspired Miura-ori honeycombs with a self-locking property},
journal = {Thin-Walled Structures},
volume = {171},
pages = {108806},
year = {2022},
issn = {0263-8231},
doi = {https://doi.org/10.1016/j.tws.2021.108806},
url = {https://www.sciencedirect.com/science/article/pii/S0263823121007795},
author = {Jianyu Gao and Zhong You}
}

@article{doi:10.1098/rspa.2023.0956,
  title={Axisymmetric blockfold origami: a non-flat-foldable Miura variant with self-locking mechanisms and enhanced stiffness},
  author={Dang, Xiangxin and Paulino, Glaucio H},
  journal={Proceedings of the Royal Society A},
  volume={480},
  number={2288},
  pages={20230956},
  year={2024},
  publisher={The Royal Society}
}

@article{Schenk2013_PNAS,
  title={Geometry of Miura-folded metamaterials},
  author={Schenk, Mark and Guest, Simon D},
  journal={Proceedings of the National Academy of Sciences},
  volume={110},
  number={9},
  pages={3276--3281},
  year={2013},
  publisher={National Academy of Sciences}
}

@article{yasuda2015reentrant,
  title = {Reentrant Origami-Based Metamaterials with Negative Poisson's Ratio and Bistability},
  author = {Yasuda, H. and Yang, J.},
  journal = {Phys. Rev. Lett.},
  volume = {114},
  issue = {18},
  pages = {185502},
  numpages = {5},
  year = {2015},
  month = {May},
  publisher = {American Physical Society},
  doi = {10.1103/PhysRevLett.114.185502},
}

@article{liu2022triclinic,
  title={Triclinic metamaterials by tristable origami with reprogrammable frustration},
  author={Liu, Ke and Pratapa, Phanisri P and Misseroni, Diego and Tachi, Tomohiro and Paulino, Glaucio H},
  journal={Advanced Materials},
  volume={34},
  number={43},
  pages={2107998},
  year={2022},
  publisher={Wiley Online Library}
}

@article{cheung2014origami,
  title={Origami interleaved tube cellular materials},
  author={Cheung, Kenneth C and Tachi, Tomohiro and Calisch, Sam and Miura, Koryo},
  journal={Smart Materials and Structures},
  volume={23},
  number={9},
  pages={094012},
  year={2014},
  publisher={IOP Publishing}
}

@article{chen2016topological,
  title={Topological mechanics of origami and kirigami},
  author={Chen, Bryan Gin-ge and Liu, Bin and Evans, Arthur A and Paulose, Jayson and Cohen, Itai and Vitelli, Vincenzo and Santangelo, Christian D},
  journal={Physical review letters},
  volume={116},
  number={13},
  pages={135501},
  year={2016},
  publisher={APS}
}

@article{tachi2012rigid,
  title={Rigid-foldable cylinders and cells},
  author={Tachi, Tomohiro and Miura, Koryo},
  journal={Journal of the international association for shell and spatial structures},
  volume={53},
  number={4},
  pages={217--226},
  year={2012},
  publisher={International Association for Shell and Spatial Structures (IASS)}
}

@misc{supp,
  note = "See Supplemental Material at
    URL-will-be-inserted-by-publisher for additional kinematic derivations and geometric details, the rigid origami model, the compression experiment setup, and Supplemental Movies S1--S3."
}

@Article{Coulais2018,
author={Coulais, Corentin
and Kettenis, Chris
and van Hecke, Martin},
title={A characteristic length scale causes anomalous size effects and boundary programmability in mechanical metamaterials},
journal={Nature Physics},
year={2018},
month={Jan},
day={01},
volume={14},
number={1},
pages={40-44},
issn={1745-2481},
doi={10.1038/nphys4269},
url={https://doi.org/10.1038/nphys4269}
}

@Article{Yang2021,
author={Yang, Hua
and M{\"u}ller, Wolfgang H.},
title={Size effects of mechanical metamaterials: a computational study based on a second-order asymptotic homogenization method},
journal={Archive of Applied Mechanics},
year={2021},
month={Mar},
day={01},
volume={91},
number={3},
pages={1037-1053},
issn={1432-0681},
doi={10.1007/s00419-020-01808-x},
url={https://doi.org/10.1007/s00419-020-01808-x}
}

@Article{Dudek2025,
author={Dudek, Krzysztof K.
and Kadic, Muamer
and Coulais, Corentin
and Bertoldi, Katia},
title={Shape-morphing metamaterials},
journal={Nature Reviews Materials},
year={2025},
month={Oct},
day={01},
volume={10},
number={10},
pages={783-798},
issn={2058-8437},
doi={10.1038/s41578-025-00828-9},
url={https://doi.org/10.1038/s41578-025-00828-9}
}

@article{wei2013geometric,
  title={Geometric mechanics of periodic pleated origami},
  author={Wei, Zhiyan Y and Guo, Zengcai V and Dudte, Levi and Liang, Haiyi Y and Mahadevan, Lakshminarayanan},
  journal={Physical review letters},
  volume={110},
  number={21},
  pages={215501},
  year={2013},
  publisher={APS}
}

@article{koryo1985method,
  title={Method of packaging and deployment of large membranes in space},
  author={Miura, Koryo},
  journal={The Institute of Space and Astronautical Science report},
  number={618},
  pages={1--9},
  year={1985},
  publisher={宇宙科学研究所}
}

@article{2019HiromiTMP,
  title={Origami-based cellular structures with in situ transition between collapsible and load-bearing configurations},
  author={Yasuda, Hiromi and Gopalarethinam, Balakumaran and Kunimine, Takahiro and Tachi, Tomohiro and Yang, Jinkyu},
  journal={Advanced Engineering Materials},
  volume={21},
  number={12},
  pages={1900562},
  year={2019},
  publisher={Wiley Online Library}
}

@article{frenzel2017three,
  title={Three-dimensional mechanical metamaterials with a twist},
  author={Frenzel, Tobias and Kadic, Muamer and Wegener, Martin},
  journal={Science},
  volume={358},
  number={6366},
  pages={1072--1074},
  year={2017},
  publisher={American Association for the Advancement of Science}
}

@article{silverberg2014using,
  title={Using origami design principles to fold reprogrammable mechanical metamaterials},
  author={Silverberg, Jesse L and Evans, Arthur A and McLeod, Lauren and Hayward, Ryan C and Hull, Thomas and Santangelo, Christian D and Cohen, Itai},
  journal={science},
  volume={345},
  number={6197},
  pages={647--650},
  year={2014},
  publisher={American Association for the Advancement of Science}
}

@article{opencv_library,
    author = {Bradski, G.},
    journal = {Dr. Dobb's Journal of Software Tools},
    title = {{The OpenCV Library}},
    year = {2000}
}

\end{document}